
%
%
\input harvmac
%
%
%
%
\ifx\answ\bigans
\else
\output={
  \almostshipout{\leftline{\vbox{\pagebody\makefootline}}}\advancepageno
}
\fi
%
%
%

%
%

%
%
\def\UCSD#1#2{\noindent#1\hfill #2%
\bigskip\supereject\global\hsize=\hsbody%
\footline={\hss\tenrm\folio\hss}}
%
%
\def\abstract#1{\centerline{\bf Abstract}\nobreak\medskip\nobreak\par #1}
%
%
%
%
\edef\tfontsize{ scaled\magstep3}
 \tfontsize  \tfontsize
 \tfontsize \font\titlei=cmmi10 \tfontsize
\font\titleis=cmmi7 \tfontsize \font\titleiss=cmmi5 \tfontsize
\font\titlesy=cmsy10 \tfontsize \font\titlesys=cmsy7 \tfontsize
\font\titlesyss=cmsy5 \tfontsize  \tfontsize
\skewchar\titlei='177 \skewchar\titleis='177 \skewchar\titleiss='177
\skewchar\titlesy='60 \skewchar\titlesys='60 \skewchar\titlesyss='60
%
%
%
%
%
\def\inv{^{\raise.15ex\hbox{${\scriptscriptstyle -}$}\kern-.05em 1}}
\def\lbar{{\lower.35ex\hbox{$\mathchar'26$}\mkern-10mu\lambda}}

%
%
%
%
\def\slash#1{\rlap{$#1$}/} 
\def\dsl{\,\raise.15ex\hbox{/}\mkern-13.5mu D} 
\def\delsl{\raise.15ex\hbox{/}\kern-.57em\partial}
\def\Ksl{\hbox{/\kern-.6000em\rm K}}
\def\Asl{\hbox{/\kern-.6500em \rm A}}
\def\Dsl{\hbox{/\kern-.6000em\rm D}} 
\def\Qsl{\hbox{/\kern-.6000em\rm Q}}
\def\gradsl{\hbox{/\kern-.6500em$\nabla$}}
%
%
\def\lspace{\ifx\answ\bigans{}\else\qquad\fi}
\def\lbspace{\ifx\answ\bigans{}\else\hskip-.2in\fi} 
%
%
\def\boxeqn#1{\vcenter{\vbox{\hrule\hbox{\vrule\kern3pt\vbox{\kern3pt
        \hbox{${\displaystyle #1}$}\kern3pt}\kern3pt\vrule}\hrule}}}
%
%
\def\mbox#1#2{\vcenter{\hrule \hbox{\vrule height#2in
\kern#1in \vrule} \hrule}}
%
%
%
%

   \def\CL{{\cal L}}
\def\CM{{\cal M}}

%
%
%
%
%
\def\del{\partial}

\def\bar#1{\overline{#1}}

\def\abs#1{\left| #1\right|}

\def\darr#1{\raise1.5ex\hbox{$\leftrightarrow$}\mkern-16.5mu #1}

%
%
\def\frac#1#2{{\textstyle{#1\over #2}}} 
%
%
%
%

\def\Tr{\mathop{\rm Tr}}

\def\GeV{{\rm GeV}}
\def\MeV{{\rm MeV}}

%
%
%
%

%
%
\def\ltap{\ \raise.3ex\hbox{$<$\kern-.75em\lower1ex\hbox{$\sim$}}\ }
\def\gtap{\ \raise.3ex\hbox{$>$\kern-.75em\lower1ex\hbox{$\sim$}}\ }
\def\gl{\ \raise.5ex\hbox{$>$}\kern-.8em\lower.5ex\hbox{$<$}\ }
\def\roughly#1{\raise.3ex\hbox{$#1$\kern-.75em\lower1ex\hbox{$\sim$}}}
%
%
\def\ie{\hbox{\it i.e.}}        
        
\def\etal{\hbox{\it et al.}}

\def\np#1#2#3{{Nucl. Phys. } B{#1} (#2) #3}
\def\pl#1#2#3{{Phys. Lett. } {#1}B (#2) #3}

\def\physrev#1#2#3{{Phys. Rev. } {#1} (#2) #3}

\relax

\def\[{\left[}
\def\]{\right]}
\def\({\left(}
\def\){\right)}

\def\dzs{D^{*0}}
\def\dz{D^0}
\def\dps{D^{*+}}
\def\dplus{D^+}
\def\dss{D_s^*}
\def\ds{D_s}
\def\chiral{$SU(3)_L\times SU(3)_R$}
\def\sproj{{1\over 2} \(1+\slash{v}\)}
\def\levi{\epsilon^{\mu\alpha\beta\lambda}}
\def\chiral{SU(3)_L\times SU(3)_R}
\def\gmu{\gamma_\mu}
\def\fpi{f_\pi}
\def\hcbar{\bar h_v^{(c)}}
\def\hc{h_v^{(c)}}
\def\[{\left[}
\def\]{\right]}
\def\gamf{\gamma_5}
\def\Br{{\rm BR}}

\noblackbox
\vskip 1.in
\centerline{{\titlefont{Radiative $D^*$ Decay Using }}}
\medskip
\centerline{{\titlefont{Heavy Quark and Chiral Symmetry}}}
\vskip .5in
\centerline{James F.~Amundson${}^{a},$ C.~Glenn Boyd${}^{a}$, Elizabeth
Jenkins${}^b$\footnote{${}^*$}{On leave from the
University of California at San Diego.}, Michael Luke${}^c$,}
\smallskip
\centerline{Aneesh V.~Manohar${}^{b\,*}$,
Jonathan L.~Rosner${}^a$, Martin J. Savage${}^c$\footnote{$^{\dagger}$}{SSC
Fellow} and
Mark B.~Wise${}^d$}
\bigskip
\centerline{\sl a) Enrico Fermi Institute and Department of Physics,
University of Chicago,}
\centerline{\sl  5640 S. Ellis Ave, Chicago, IL 60637}
\centerline{\sl b) CERN TH Division, CH-1211 Geneva 23, Switzerland}
\centerline{\sl c) Department of Physics, University of California at San
Diego,}\centerline{\sl 9500 Gilman Drive, La Jolla, CA 92093}
\centerline{\sl d) California Institute of Technology, Pasadena, CA 91125}
\vfill
\abstract{The implications of chiral $\chiral$ symmetry and heavy
quark symmetry for the radiative decays $\dzs\to\dz\gamma$, $\dps\to
\dplus\gamma$, and $\dss\to\ds\gamma$ are discussed.  Particular attention is
paid to $SU(3)$ violating contributions of order $m_q^{1/2}$.
Experimental data on these radiative decays provide constraints on the
$D^* D\pi$ coupling.
}
\vfill
\UCSD{\vbox{
\hbox{UCSD/PTH 92-31}
\hbox{CALT-68-1816}
\hbox{EFI-92-45}
\hbox{CERN-TH.6650/92}
\hbox{hep-ph@xxx/9209241}}}{September 1992}
\eject

Recent CLEO data \ref\cleo{The CLEO Collaboration (F.~Butler, \etal),
CLNS-92-1143 (July 1992)}
(see Table 1) have brought the $\dzs$ and $\dps$ branching
ratios into agreement with expectations based on the constituent quark
model \ref\cqm{J.~L.~Rosner, in {\it Particles and Fields 3, Proceedings
of the Banff Summer Institute, Banff Canada 1988}, A.~N.~Kamal and
F.~C.~Khanna, eds., World Scientific, Singapore (1989), p. 395\semi
L.~Angelos and G.~P.~Lepage, \physrev{D45}{1992}{3021}}.
In this letter, the rates for
$D^*$ decay are described
in a model independent framework which incorporates the constraints on the
decay amplitudes imposed by the heavy quark and
chiral $\chiral$ symmetries of QCD.
\bigskip
\input tables
\centerline{{\bf Table 1: $D^*$ Branching Ratios (\%)}}
\bigskip
\begintable
Decay Mode |  ~~Branching Ratio~~\cr
$\dzs\to\dz\pi^0$ | $63.6\pm2.3\pm3.3$\crnorule
$\dzs\to\dz\gamma$ | $36.4\pm2.3\pm3.3$\cr
$\dps\to\dz\pi^+$ |$ 68.1\pm1.0\pm1.3$\crnorule
$\dps\to\dplus\pi^0$ | $30.8\pm0.4\pm0.8$\crnorule
$\dps\to\dplus\gamma$ | $1.1\pm1.4\pm1.6$
\endtable
\bigskip

At low momentum the strong interactions of the $D$ and $D^*$ mesons are
described by the chiral Lagrange density
\ref\wbdy{M.~Wise, \physrev{D45}{1992}{2188}\semi
G.~Burdman and J.~F.~Donoghue, \pl{280}{1992}{287}\semi
T.~M.~Yan \etal, \physrev{D46}{1992}{1148}}
\eqn\lag{\eqalign{\CL&=-i\Tr\bar H_a v_\mu \del^\mu H_a
    +\frac{i}{2}\Tr\bar H_a H_b v_\mu\[\xi^\dagger\del^\mu\xi+
    \xi\del^\mu\xi^\dagger\]_{ba}\cr
    &+\frac{i}{2}g\Tr\bar H_a H_b\,\gamma_\mu\gamma_5
    \[\xi^\dagger\del^\mu\xi-\xi\del^\mu\xi^\dagger\]_{ba}+\cdots}}
where the ellipsis denotes operators suppressed by factors of $1/m_Q$
and operators with more derivatives
or factors of the light quark mass matrix.  In Eq.~\lag, $v^\mu$ is the four
velocity of the heavy meson.  The field $\xi$ is written in terms of
the octet of pseudo-Nambu-Goldstone bosons
\eqn\pseudo{\xi=\exp\(i\CM/f\),}
where
\eqn\cm{\CM=\pmatrix{{\textstyle{1\over\sqrt{2}}}\pi^0+{\textstyle
    {1\over\sqrt{6}}}\eta&\pi^+&K^+\cr\pi^-
    &{\textstyle{-{1\over\sqrt{2}}}}\pi^0+
    {\textstyle{1\over\sqrt{6}}}\eta&K^0\cr K^-&\bar K^0&-{\textstyle\sqrt
    {2\over 3}}\eta}.}
At tree level $f$ can be set equal to $f_\pi$, $f_K$ or $f_\eta$.  Our
normalization convention has $f_\pi\simeq 132\,\MeV$.
Under chiral $\chiral$ transformations,
\eqn\transf{\xi\to L\xi U^\dagger = U\xi R^\dagger,}
where $L\in SU(3)_L$ and $R \in SU(3)_R$, and $U$ is defined implicitly
by Eq.~\transf.
$H_a$ is a $4 \times 4$ matrix that contains the $D$ and $D^*$ fields:
\eqn\defineh{\eqalign{H_a&=\sproj\[D^{*\mu}_a\gmu - D_a\gamf\],\cr
             \bar H_a&=\gamma^0 H_a^\dagger \gamma^0\,.}}
The index $a$ represents light quark flavor, where $(D_1,D_2,D_3)=
(\dz,\dplus,\ds)$ and $(D_1^*,D_2^*,D_3^*)=(\dzs,\dps,\dss)$.  Under $SU(2)_v$
heavy quark spin symmetry and chiral $\chiral$ symmetry, $H_a$ transforms
as
\eqn\trans{H_a\to S(HU^\dagger)_a\,,}
where $S\in SU(2)_v$.
The $D^*D\pi$ coupling constant $g$ is responsible for the $D^*\to D\pi$
decays.  At tree
level,
\eqn\width{\Gamma(\dps\to\dz\pi^+)={g^2\over
6\pi\fpi^2}\abs{\vec{p_\pi}}^3\,.}
The decay width for $\dps\to\dplus\pi^0$
is a factor of two smaller by isospin symmetry.
The experimental upper limit \ref\accmor{The ACCMOR
Collaboration (S.~Barlag \etal), \pl{278}{1992}{480}}\ on
the $\dps$ width of 131 keV
when combined with the $\dps\to\dplus\pi^0$ and $\dps\to\dz\pi^+$ branching
ratios in Table 1 leads to the limit $g^2\ltap 0.5$.

The axial vector current obtained from the Lagrangian \lag\ is
\eqn\ax{\bar q_a\, T^A_{ab}\,\gamma_\nu\gamf\, q_b=-g\,\Tr\bar H_a H_b
\,\gamma_\nu\gamf\, T^A_{ba}+\cdots\,.}
In Eq.~\ax\ the ellipsis represents terms containing one or more Goldstone
boson fields and $T^A$ is a flavor $SU(3)$ generator.  Treating the
quark fields in Eq.~\ax\ as constituent quarks and using the
nonrelativistic quark model to estimate the $D^*$ matrix element of the
l.h.s.~of Eq.~\ax\ gives $g=1$.  (A similar estimate of the pion-nucleon
coupling gives $g_A=5/3$.)  In the chiral quark model
\ref\mg{A.~V.~Manohar and H.~Georgi, \np{234}{1984}{189}}\ there
is a
constituent quark-pion coupling.  Using the measured pion-nucleon
coupling to determine the constituent quark pion coupling gives $g\simeq
0.8$.  Thus various constituent quark model estimates lead to the
expectation that $g$ is near unity. In this paper, however, we wish to adopt a
model independent approach to radiative $D^*$ decay.
{}From the point of view of
chiral perturbation theory $g$ is a free parameter and its value must be
determined from experiment.

The $D_a^*\to D_a\gamma$ matrix element has the form
\eqn\rad{\CM(D_a^*\to D_a\gamma)=e\mu_a\ \levi\,\epsilon_\mu^*(\gamma)
\,v_\alpha\, k_\beta\, \epsilon_\lambda(D^*),}
where $e\mu_a/2$ is the transition magnetic moment,
$k$ is the photon momentum, $\epsilon(\gamma)$ is the polarization
of the photon and $\epsilon(D^*)$ is the polarization of the $D^*$.  The
resulting decay rate is
\eqn\raddec{\Gamma(D_a^*\to D_a\gamma)={\alpha\over 3}\abs{\mu_a}^2
\vert\vec{k}\vert^3\,.}
The $D_a^*\to D_a\gamma$ matrix element gets contributions from the
photon coupling to the light quark part of the electromagnetic current,
$\frac23\,\bar u\gmu u-\frac13\, \bar d\gmu d-\frac13\,
\bar s\gmu s$, and the photon
coupling to the heavy charm quark part of the electromagnetic current,
$\frac23\, \bar c \gmu c$.  The part of $\mu_a$ that comes from the charm
quark piece of the electromagnetic current, $\mu^{(h)}$, is determined
by heavy quark symmetry.  In the effective heavy quark theory \ref\hqet
{E.~Eichten and B.~Hill, \pl{234}{1990}{511}\semi
H.~Georgi, \pl{240}{1990}{447}\semi
A.~F.~Falk, B.~Grinstein and M.~Luke, \np{357}{1991}{185}},
the Lagrange density for strong and electromagnetic interactions of
the charm quark is
\eqn\hqlag{\eqalign{\CL=&\hcbar \left(iv\cdot D\right) \,\hc +{1\over
2m_c}\hcbar(iD)^2\hc\cr
&-{g_s\over 2m_c}\hcbar\sigma^{\mu\nu}T^a\hc G_{\mu\nu}^a
-{e\over 3m_c}\hcbar\sigma^{\mu\nu}\hc F_{\mu\nu}+\cdots\,.}}
In Eq.~\hqlag, $D_\mu$ is the covariant derivative
\eqn\cov{D_\mu=\del_\mu+ig_s A_\mu^a T^a+\frac23 ieA_\mu,}
where $g_s$ is the strong coupling and $e$ is the electromagnetic coupling.
The ellipsis denotes terms with
more factors of $1/m_c$.  It is to be understood that the
operators and couplings in Eq.~\hqlag\ are evaluated at a subtraction
point $\mu=m_c$, and that corrections of order $\alpha_s(m_c)$ have been
neglected.  The last term in Eq.~\hqlag\ is responsible for a $D^*$ to $D$
transition matrix element $\mu^{(h)}$.
By
heavy quark symmetry \ref\hqsym{N.~Isgur and M.~B.~Wise,
\pl{232}{1989}{113}; \pl {237}{1990}{527}},
\eqn\mua{\mu^{(h)}={2\over 3 m_c}\,,}
where $\mu^{(h)}$ is
independent of the light quark flavor.
Perturbative $\alpha_s(m_c)$ corrections to the above are computable,
while corrections suppressed by a power of $1/m_c$ are related to those
which occur in semileptonic $\bar B\to D^*e\bar\nu_e$ decays
\ref\mel{M.~Luke, \pl{252}{1990}{447}}.  At order
$1/m_c^2$, Eq.~\mua\ becomes $\mu^{(h)}=(2/3 m_c)\[1-4\xi_+(1)/m_c\]$,
where $\xi_+$ is defined in Ref.~\mel.

The part of $\mu_a$ that comes from the photon coupling to the light
quark piece of the electromagnetic current, $\mu_a^{(\ell)}$, is not
fixed by heavy quark symmetry.  The light quark piece of
the electromagnetic current transforms as an octet under $SU(3)$
flavor symmetry. Since there is only one way to combine an 8, 3 and $\bar
3$ into a singlet, in the limit of $SU(3)$ symmetry, the $\mu_a^{(\ell)}$
are expressible in terms of a single reduced matrix element,
\eqn\mul{\mu_a^{(\ell)}=Q_a\beta\,,}
where $\beta$ is an unknown constant and
$Q_a$ denotes the light quark charges $Q_1=2/3, Q_2=-1/3,
Q_3=-1/3$.  In the nonrelativistic constituent quark model $\beta\simeq
3\,\GeV^{-1}$.  Note that Eq.~\mul\ includes effects suppressed by powers
of $1/m_c$, since it follows from using only $SU(3)$ symmetry.

The leading $SU(3)$-violating contribution to the transition amplitudes
has a nonanalytic dependence on $m_q$
of the form $m_q^{1/2}$ which arises from
the one-loop Feynman diagrams shown in \fig\diag{Diagrams giving the
leading
non-analytic contributions to $\mu_a^{(\ell)}$.}. The
strange quark mass, $m_s$, is not very small, and so the corrections to
Eq.~\mul\ from $SU(3)$ violation may be comparable to $\mu^{(h)}$,
which is suppressed by $1/m_c$ relative to $\mu^{(\ell)}$.
Including the
leading SU(3) violations,
$\mu_a^{(\ell)}$ becomes
\eqn\nonan{\eqalign{&\mu_1^{(\ell)}={2\over 3}\beta-{g^2 m_K\over
4\pi f_K^2}-{g^2 m_\pi\over 4\pi f_\pi^2},\cr
&\mu_2^{(\ell)}=-{1\over 3}\beta+{g^2 m_\pi\over 4\pi f_\pi^2},\cr
&\mu_3^{(\ell)}=-{1\over 3}\beta+{g^2 m_K\over4\pi f_K^2}\,.}}
The difference between using $f=\fpi$ and $f=f_K$ in Eq.~\nonan\ is
a higher order
effect. We have chosen to use $f=f_K\simeq 1.22\,\fpi$ for loops involving
kaons and $f=f_\pi$ for loops involving pions.
For $m_K\ne m_\pi$, the one loop contribution to $\mu_1^{(\ell)}$,
$\mu_2^{(\ell)}$ and $\mu_3^{(\ell)}$ is not in the ratio $2:-1:-1$ and
hence violates $SU(3)$.
It is easy to understand
why the one-loop correction proportional to $m_K$ is different
for the $\dzs\to\dz\gamma$
and $\dps\to\dplus\gamma$ decays. Strong interactions can change
a $\dzs$ into a virtual $K^-\dss$ pair,
while the $\dps$ changes into a virtual $K^0\dss$ pair.  In the latter
case the virtual kaon is neutral and doesn't couple to the photon. Thus
there is no $m_s^{1/2}$ correction to $\mu_2^{(\ell)}$. The most
important correction to Eq.~\nonan\ comes from $SU(3)$
violating terms of order $m_s$. These terms are analytic in the strange
quark mass, and are not determined by the lowest order Lagrangian.

Using
\eqn\summu{\mu_a=\mu_a^{(\ell)}+\mu^{(h)},}
with $\mu_a^{(\ell)}$ and $\mu^{(h)}$ given by Eqs.~\nonan\ and
\mua\ respectively, determines the rates for $\dzs\to\dz\gamma$,
$\dps\to\dplus\gamma$
and $\dss\to\ds\gamma$ in terms of $\beta$ and $g$.  Combining
this with Eq.~\width\ and using the measured value of
$\Br(\dzs\to\dz\gamma)/\Br(\dzs\to\dz\pi^0)$ gives $g$  as a
function of the branching ratio for $\dps\to\dplus\gamma$.
This in fact gives four different solutions for $g^2$; we eliminated
three of these by imposing the constraints $g<1$ (as required by
Ref.~\accmor) and
$\mu_a^{(\ell)}>\mu^{(h)}$ \ie, the light quark transition moment
is greater than that of the heavy quark.
The result is shown
in \fig\branch{The coupling constant $g$ as a function of
$\Br(\dps\to\dplus\gamma)$ including leading
SU(3)-breaking effects.  The shaded region indicates the uncertainty due
to the $1\sigma$ variations in $\Br(\dzs\to\dz\pi^0)$ and
$\Br(\dzs\to\dz\gamma)$.
The arrows indicate the 90\% confidence level
limits on $\Br(\dps\to\dplus\gamma)$ and the $\dps$ width.}.
(We have taken $m_c=1.7\,\GeV$.)
Note that the favored values for
$g$ are smaller than what is
expected on the basis of the nonrelativistic constituent quark model.
Since $1/m_c$ effects have been included in the radiative $D^*$ decays,
the value of $g$ extracted in this way is an ``effective'' value of $g$
that includes $1/m_c$ corrections.
{}From Eq.~\width\ and our values of $g$ we can compute the total width
of the $\dps$ as a function of $\Br(\dps\to\dplus\gamma)$; this is plotted
in \fig\totalwidth{Width of the $\dps$ as a function of
$\Br(\dps\to\dplus\gamma)$ including leading
SU(3)-breaking effects.  The shaded region indicates the uncertainty due
to the $1\sigma$ variations in $\Br(\dzs\to\dz\pi^0)$ and
$\Br(\dzs\to\dz\gamma)$.
The arrows indicate the 90\% confidence level
limits on $\Br(\dps\to\dplus\gamma)$ and the $\dps$ width.}.

The $SU(3)$ violation plays an important role in our analysis.
Fig.~\the\figno\nfig\relcont{Relative contributions to $\mu_1$
of $\mu^{(h)}$
(dashed-dotted
line), $\beta$ (dotted line), and the one-loop nonanalytic $m_q^{1/2}$
term
(solid line) to the matrix element for
$\dzs\to\dz\gamma$.}\ shows the absolute values of the
relative contributions to $\mu_1$ of $\mu^{(h)}$
(dashed-dotted line), $\beta$
(dotted line) and the one-loop nonanalytic contribution
to $\mu_1^{(\ell)}$ (solid line).
The values have all been multiplied by 3/2, so that the dotted line
is normalized to $\beta$.
Note that values of $\beta$ near the non-relativistic
constituent quark model expectation of $\approx 3$ $\GeV^{-1}$
favor a small $\dps\to\dplus\gamma$ branching ratio,
and hence smaller values of $g$.
In \fig\ignore{The coupling constant $g$ as a function of
$\Br(\dps\to\dplus\gamma)$ ignoring $SU(3)$ violation.
The shaded region indicates the uncertainty due
to the $1\sigma$ variations in $\Br(\dzs\to\dz\pi^0)$ and
$\Br(\dzs\to\dz\gamma)$.  The arrows indicate the 90\% confidence level
limits on $\Br(\dps\to\dplus\gamma)$ and the $\dps$ width.}
the value of $g$ that follows from
neglecting $SU(3)$ violation (\ie\ using Eq.~\mul\ for $\mu_a^{(\ell)}$)
is shown.  Larger values of $g$ are favored when $SU(3)$ violation is
neglected.

Nonanalytic dependence on $m_s$
similar to what we have found in radiative $D^*$ decay
occurs in the $\ds-\dplus$ mass difference.
Including effects up to order $m_s^{3/2}$ \ref\goity{J.~L.~Goity,
CEBAF-TH-92-16 (1992)}
\eqn\massdiff{m_{\ds}-m_{\dplus}=C m_s-{3g^2\over 64\pi f_K^2}\(2
m_K^3+m_\eta^3\),}
where we have set $m_u=m_d=0$ and $C$ is an unknown constant.
Experimentally, $m_{\ds}-m_{\dplus}\simeq\,100\ \MeV$.  The magnitude of the
nonanalytic part is about 50\% of the mass difference for
$g=0.5$.  This gives us some confidence that the expansion is well
behaved for at least some of the range of $g$'s in \branch.

The analysis in this paper allows us to predict the $\dss\to\ds\gamma$
rate as a function of the $\dps\to\dplus\gamma$ branching ratio.
However, for $\dss\to\ds\gamma$ there is a strong cancellation
between $\mu_3^{(\ell)}$ and $\mu^{(h)}$, resulting in a very small
$\dss$ width.  (Note that $\dss\to\ds\pi$ is forbidden by isospin.)
In this situation, $SU(3)$ violating terms of order $m_s$ may be
very important.

Since heavy quark symmetry ensures that $g$ and $\beta$ are
the same in the $b$ and $c$ systems (up to corrections of
order $1/m_c$), the results of this paper can be used to predict
the widths for radiative $B^*$ decay.    Neglecting effects of order
$1/m_b$ and $1/m_c$,
Eq.~\raddec\ becomes
\eqn\bwidth{\Gamma\(B^*_a\to B_a\gamma\)={\alpha\over 3}\vert
\mu_a^{(\ell)}\vert^2\vert\vec{k}\vert^3}
where $\mu_a^{(\ell)}$ is given by Eq.~\nonan.
An analysis of the radiative decays of charmed baryons using the
same methods is possible.  Unfortunately, at the present time there
is no experimental information on radiative charmed baryon decays.

Work similar to that presented in this paper has also been done
by Cho and
Georgi \ref\chgeo{P.~Cho and H.~Georgi, Harvard Preprint HUTP-02/A043}.  We are
grateful to them for communicating their results to us prior to
publication. This work was supported in part by the Department of Energy
under grant number DOE-FG03-90ER40546 and contract number
DEAC-03-81ER40050, and by a National Science Foundation Presidential
Young Investigator award number PHY-8958081.  MJS
acknowledges the support of a Superconducting Supercollider
National Fellowship from the Texas National Research Laboratory
Commission under grant FCFY9219.

\listrefs
\listfigs

\bye